\documentclass[%
superscriptaddress,
showpacs,preprintnumbers,
amsmath,amssymb,
aps,
twocolumn
]{revtex4-1}
\usepackage{ulem}
 \usepackage{graphicx}
\usepackage{dcolumn}
\usepackage{bm}
\usepackage{color}
\usepackage{mathcomp}

\begin {document}

\title{Multiple binding modes of AKT on PIP$_3$-containing membranes}

\author{Yuki Nakagaki}
\affiliation{Department of System Design Engineering, Keio University, Yokohama, Kanagawa 223-8522, Japan}

\author{Eiji~Yamamoto}
\email{eiji.yamamoto@sd.keio.ac.jp}
\affiliation{Department of System Design Engineering, Keio University, Yokohama, Kanagawa 223-8522, Japan}


\begin{abstract}
The PI3K/AKT signaling pathway is triggered by recruitment of AKT to cellular membranes.
Although AKT is a multidomain serine/threonine kinase composed of an N-terminal pleckstrin homology (PH) domain and a C-terminal kinase domain, how these domains cooperate to regulate AKT activation on membranes remains unclear at the molecular level.
Here, using molecular dynamics simulations of full-length AKT on PIP$_3$-containing lipid bilayers, we identify four distinct membrane-binding modes that differ in the orientations and membrane contacts of the PH and kinase domains.
In addition to PIP$_3$ binding to the PH domain, we observe specific PIP$_3$ interactions with basic residues in the kinase domain.
In the most stable mode, PIP$_3$ interacts with both the canonical and a secondary binding site in the PH domain, while the kinase domain adopts an orientation in which the activation-loop phosphorylation site is exposed to the solvent.
Interestingly, the populations of these binding modes depend on the PIP$_3$ concentration in the membrane, leading to changes in the preferred orientation of AKT.
These findings shed light on how lipid recognition by the PH domain and the kinase domain of AKT cooperatively shape its membrane-bound conformations.
\end{abstract}

\maketitle

Cell signaling and trafficking frequently involve the reversible association of peripheral membrane proteins (PMPs) with cell membranes~\cite{Cho2005}.
AKT is a PMP that belongs to the AGC (protein kinase A, G, and C) family of serine/threonine kinases~\cite{hanks1995eukaryotic} and is a central component of the PI3K/AKT signaling pathway, which regulates cell growth, metabolism, differentiation and protein synthesis~\cite{Siess2019}.
Hyperactivation of AKT is frequently observed in human cancers and tissue overgrowth disorders~\cite{Carpten2007, Lindhurst2011, Nellist2015, Lee2012}, whereas inactivation of AKT2 leads to insulin resistance and diabetes~\cite{George2004}.
AKT activation occurs downstream of phosphoinositide 3-kinase (PI3K)~\cite{franke1995protein}.
In response to extracellular signals such as growth factors or cytokines, receptor tyrosine kinases or G protein-coupled receptors activate PI3K~\cite{Manning2017}.
Activated PI3K catalyzes the production of phosphatidylinositol (3,4,5)-trisphosphate (PI(3,4,5)P$_3$, hereafter PIP$_3$) from membrane phospholipids.

AKT is a multidomain serine/threonine kinase composed of an N-terminal pleckstrin homology (PH) domain, a flexible linker, and a C-terminal kinase domain(Fig.~\ref{Distance}A).
The PH domain specifically recognizes phosphoinositides such as PIP$_3$ and phosphatidylinositol (3,4)-bisphosphate (PIP$_2$) and is known to anchor AKT to the membrane~\cite{james1996specific, Bellacosa1998}.
In the cytosol, AKT adopts an autoinhibited ``PH-in'' conformation, in which the PH domain interacts with the kinase domain and masks the activation sites~\cite{Lucic2018}.
Upon recruitment to PIP$_3$-enriched membranes, binding of the PH domain to PIP$_3$ induces conformational changes in AKT on the membrane~\cite{Calleja2007, Lucic2018, Truebestein2021, Ebner2017}.
Subsequently, AKT is phosphorylated at Thr308 by PDK1~\cite{Alessi1997} and at Ser473 by mTORC2~\cite{Sarbassov2005} on the membrane, leading to full activation.
Once activated, AKT can dissociate from the membrane and phosphorylate a wide range of substrate proteins in the cytosol, nucleus, or on the membrane~\cite{Manning2017}, thereby promoting cell survival and other downstream processes.
Therefore, revealing the molecular mechanisms of PI3K/AKT signaling is essential for understanding the pathogenesis of these diseases.
Furthermore, AKT serves as a critical signaling hub that integrates oncogenic signals from both extracellular and intracellular sources, making it an attractive target for therapeutic intervention.
Consequently, numerous studies have been conducted in the field of drug discovery~\cite{Cheng2005, GarciaEcheverria2008, He2021}.

Several experimental and computational studies have investigated how the PH domain of AKT interacts with PIP$_3$ in lipid bilayers~\cite{james1996specific, Bellacosa1998, Milburn2003, Yamamoto2016, LeHuray2022}.
A recent study suggests that both canonical and secondary  binding sites within the PH domain of AKT contribute to membrane binding and orientation stabilization~\cite{Soteriou2025}.
However, these studies mainly focus on the isolated PH domain or do not fully resolve how full-length AKT behaves on membranes.
In particular, the membrane-bound dynamics of full-length AKT, including the relative orientation and cooperative motions of the PH and kinase domains and their interactions with PIP$_3$, remain unclear at the molecular level.

Here, we use coarse-grained molecular dynamics (CG-MD) simulations to investigate the membrane-bound dynamics of full-length AKT on PIP$_3$-containing lipid bilayers.
By quantifying the orientations and membrane contacts of the PH and kinase domains, we identify four distinct membrane-binding modes and characterize the local PIP$_3$ density and residue-level PIP$_3$ interactions associated with each mode.
These results shed light on how lipid recognition by the PH domain and PIP$_3$ interactions of the kinase domain cooperatively shape the membrane-bound conformations of AKT during PI3K/AKT signaling.

\section*{Results}
\subsection*{AKT binds to the lipid bilayer via its PH domain}
\begin{figure}[tb]
\includegraphics[width = 80 mm,bb= 0 0 406 611]{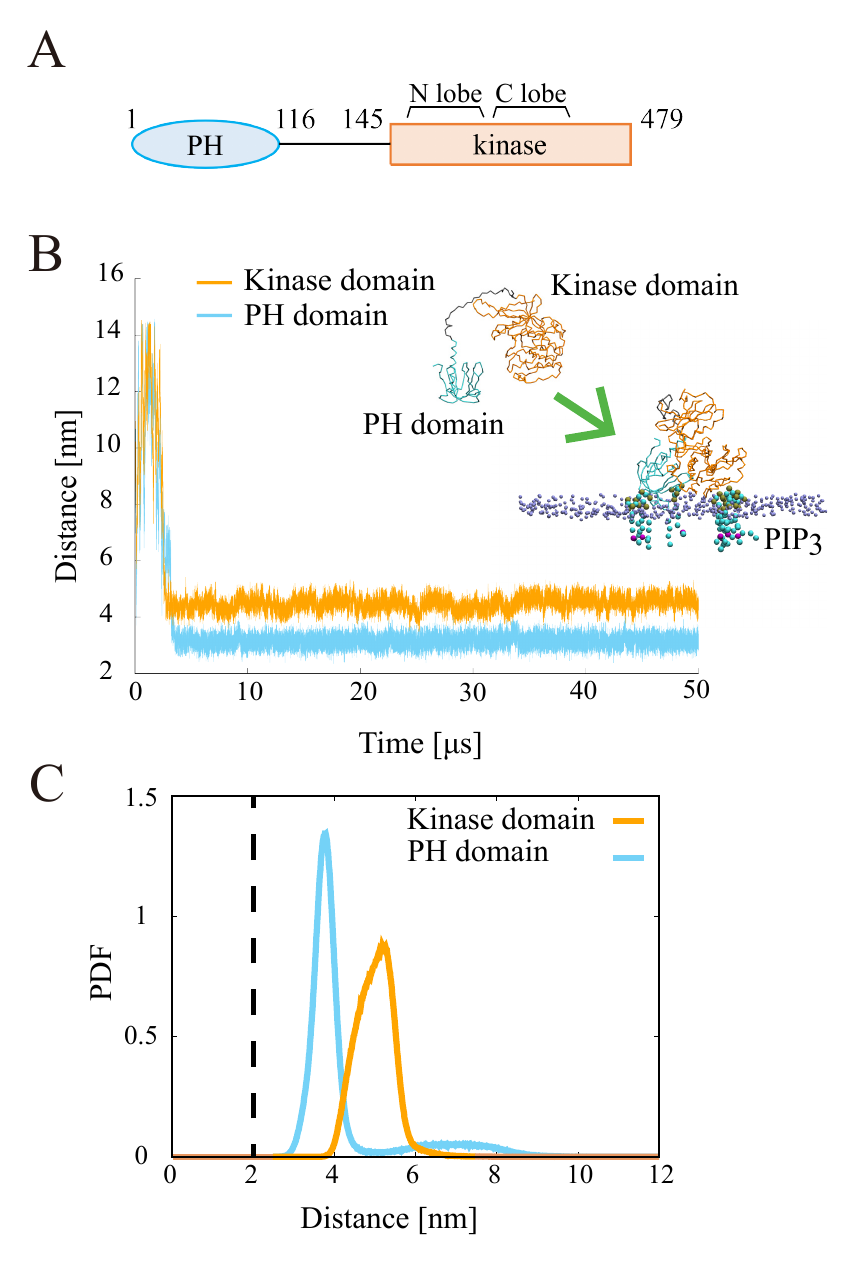}
\caption{AKT binds to the lipid bilayer via its PH domain.
(A)~Schematic representation of AKT.
(B)~Distance between the COM of each domain and the lipid bilayer as a function of simulation time.
The inset shows a snapshot from a representative simulation illustrating membrane-bound AKT.
The PH domain and kinase domain are shown in light blue and orange, respectively.
PIP$_3$ lipids in the membrane are highlighted.
(C)~Probability density functions of the distance between the COM of each domain and the lipid bilayer.
The dashed line indicates the membrane surface (phosphate group).}
\label{Distance}
\end{figure}

In ten independent CG-MD simulations of $50\,\mu\mathrm{s}$ each, full-length AKT was initially placed approximately $15\,\mathrm{nm}$ above the center of mass (COM) of a PIP$_3$-containing lipid bilayer.
By tracking the time variation of the $z$ component of the distance between the COM of each domain (PH or kinase) and that of the lipid bilayer, we observed that AKT first diffuses freely in the aqueous environment and then begins to interact continuously with the lipid bilayer after approximately $2\,\mu\mathrm{s}$ (Fig.~\ref{Distance}B).
This behavior was consistently observed in all ten independent simulations (Fig.~S\ref{support2}).

To further characterize the association process of AKT with the membrane, we calculated the probability density functions (PDFs) of the COM distances along the $z$ axis (Fig.~\ref{Distance}C).
For the PH domain, we observed a strong peak at $3.8\,\mathrm{nm}$ and a smaller peak at $7.0\,\mathrm{nm}$.
For the kinase domain, a peak appeared at $5.2\,\mathrm{nm}$.
The $3.8\,\mathrm{nm}$ peak of the PH domain corresponds to a membrane-bound state, whereas the $7.0\,\mathrm{nm}$ peak suggests configurations in which the kinase domain interacts with the membrane while the PH domain remains in the solvent.
These findings indicate that, under our simulation conditions, AKT predominantly interacts with the lipid bilayer via its PH domain, consistent with its known lipid-binding function~\cite{james1996specific, Bellacosa1998, Milburn2003, Yamamoto2016, LeHuray2022, Soteriou2025}.

\begin{figure*}[tb]
\centering
\includegraphics[width = 150 mm,bb= 0 0  616  620]{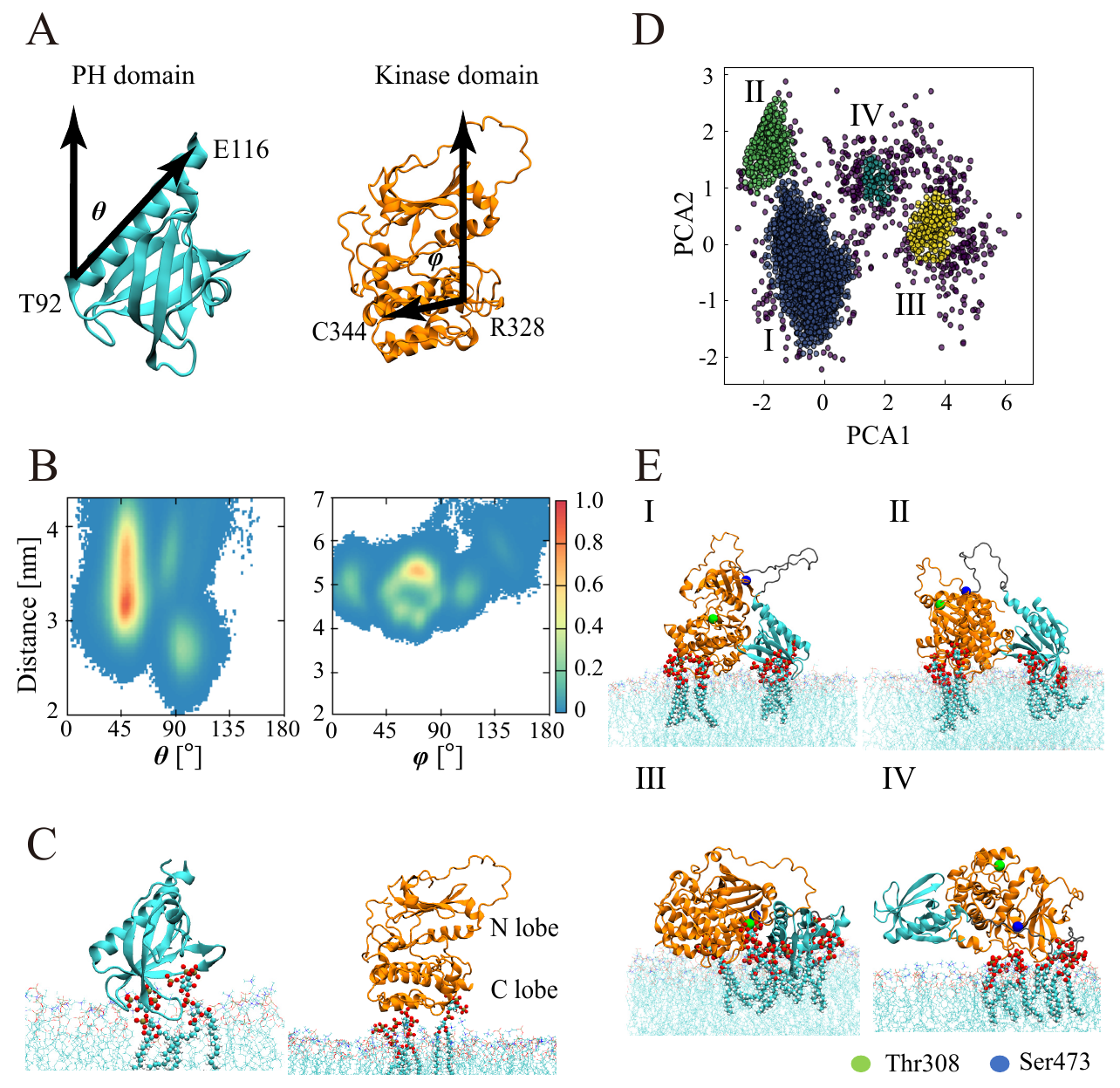}
\caption{Membrane-binding modes and orientations of AKT on the lipid bilayer.
(A)~Definition of the orientation angles.
$\theta$ is defined as the angle between the bilayer normal and the $\alpha$-helix (T92--E116) in the PH domain (left), and $\phi$ is defined as the angle between the bilayer normal and the $\alpha$-helix (R328--C344) in the kinase domain (right).
(B)~Normalized density maps for the PH domain (left) and the kinase domain (right) shown as a function of the orientation angle and the $z$-component of the domain COM--bilayer distance.
(C)~Representative orientations of the PH domain (left) and the kinase domain (right). 
PIP$_3$ lipids in the membrane are highlighted, with their oxygen atoms shown in red.
(D)~PCA analysis of the AKT orientation.
(E)~Four binding modes identified by DBSCAN clustering in the PCA space.}
\label{orientation}
\end{figure*}

\begin{figure*}[tb]
\centering
\includegraphics[width = 160 mm,bb= 0 0  649  457]{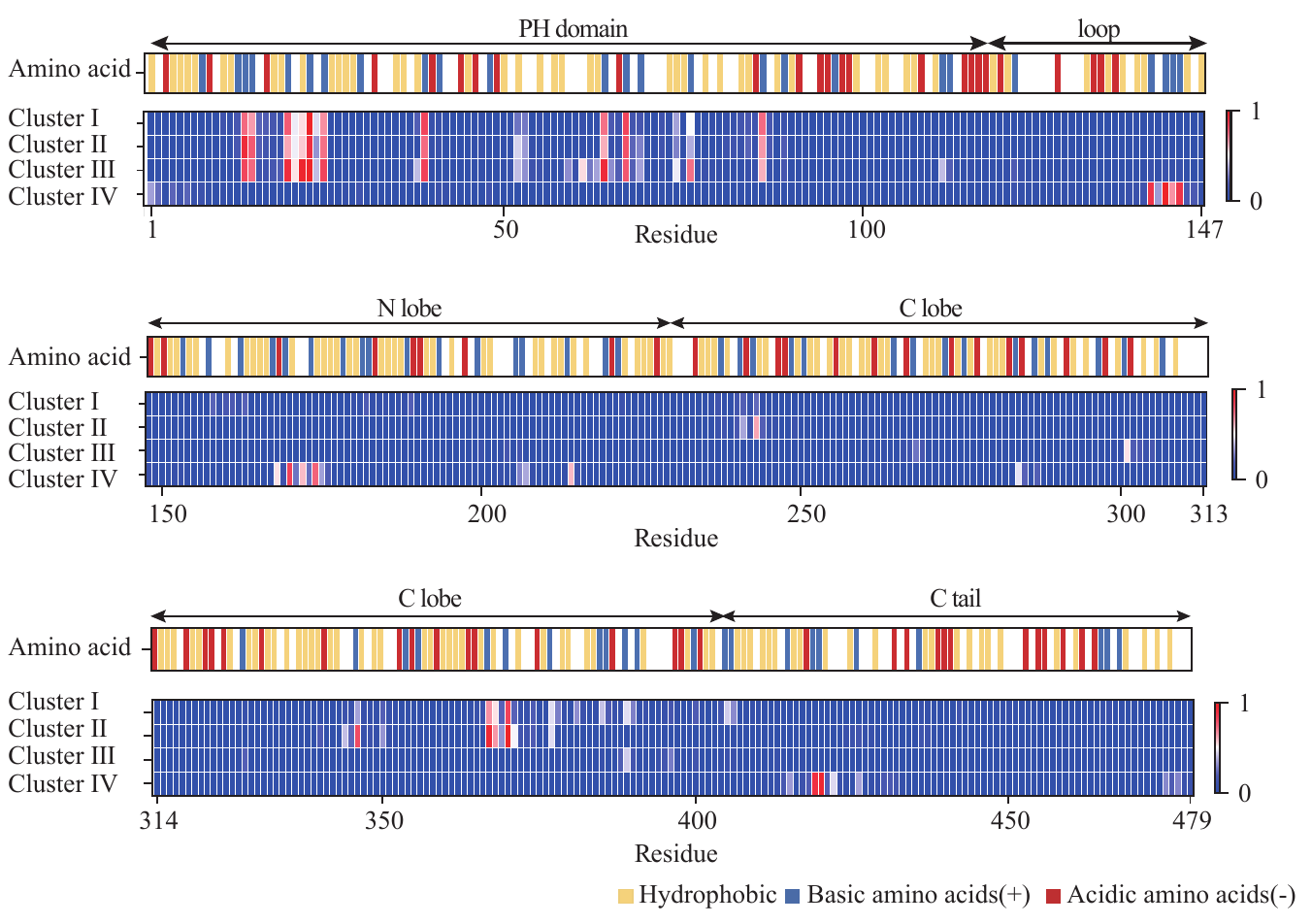}
\caption{Residues critical for PIP$_3$ binding across the four binding modes.
Normalized contacts were calculated using data from each cluster. For normalization, the number of contacts between a given residue and phosphate headgroups  was divided by the maximum number of contacts observed within each cluster.
Hydrophobic, basic, acidic, and other residues are shown in yellow, blue, red, and white, respectively.}
\label{contact}
\end{figure*}

\begin{figure*}[tb]
\centering
\includegraphics[width = 170 mm,bb= 0 0  890 809]{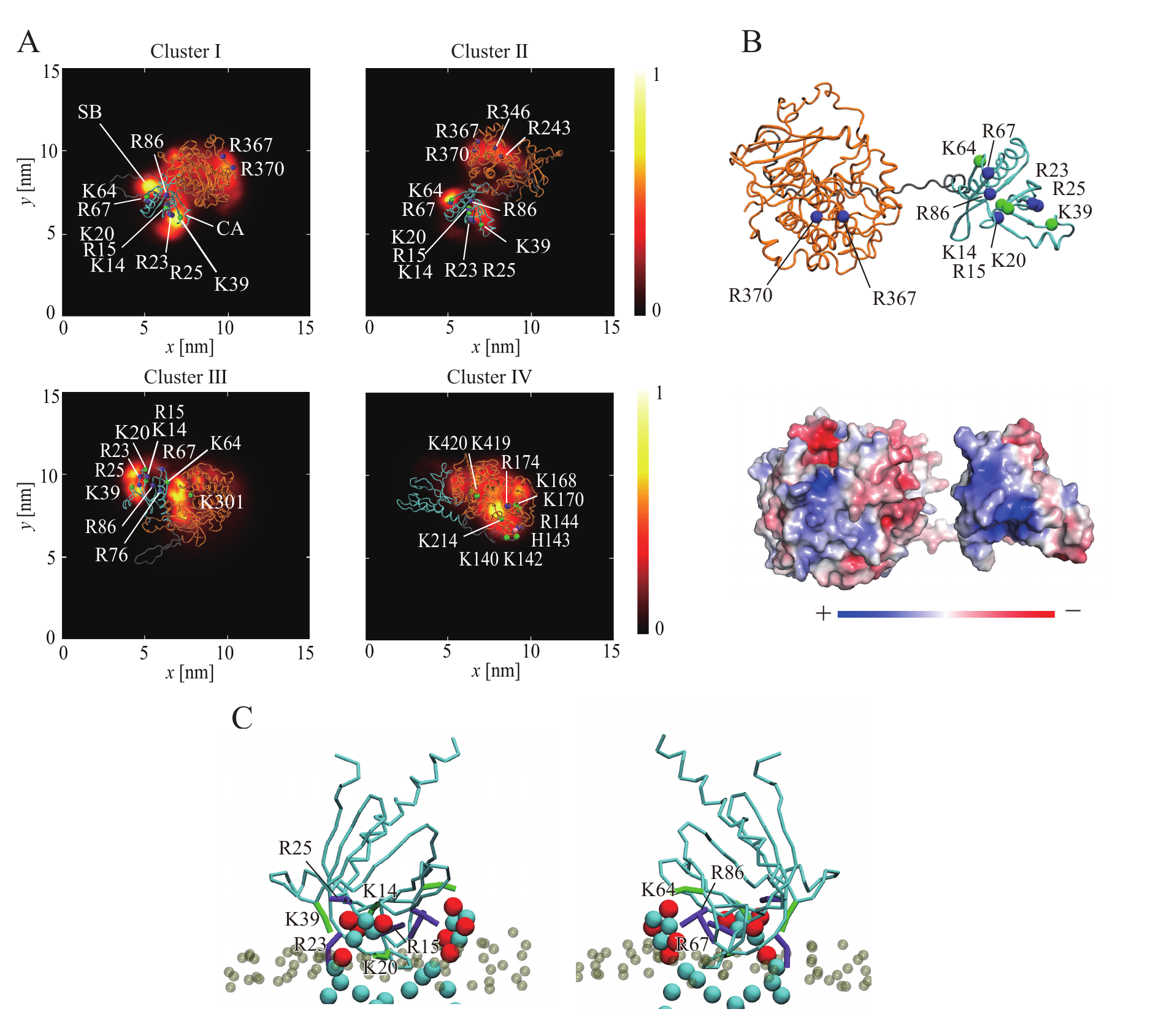}
\caption{Multiple PIP$_3$ binding sites on AKT.
(A)~Normalized density maps of PIP$_3$ phosphate headgroups in the lipid bilayer for each binding mode.
Arg (purple) and Lys (green) residues with high contact frequency with PIP$_3$ are shown as spheres. CA and SB represent the canonical and secondary binding sites, respectively.
(B)~Electrostatic potential map of AKT, highlighting the surface charge distribution.
(C)~Key residues in Cluster I involved in PIP$_3$ association.}
\label{PIP_density}
\end{figure*}

\begin{figure*}[tb]
\centering
\includegraphics[width = 150 mm, bb= 0 0 766 642]{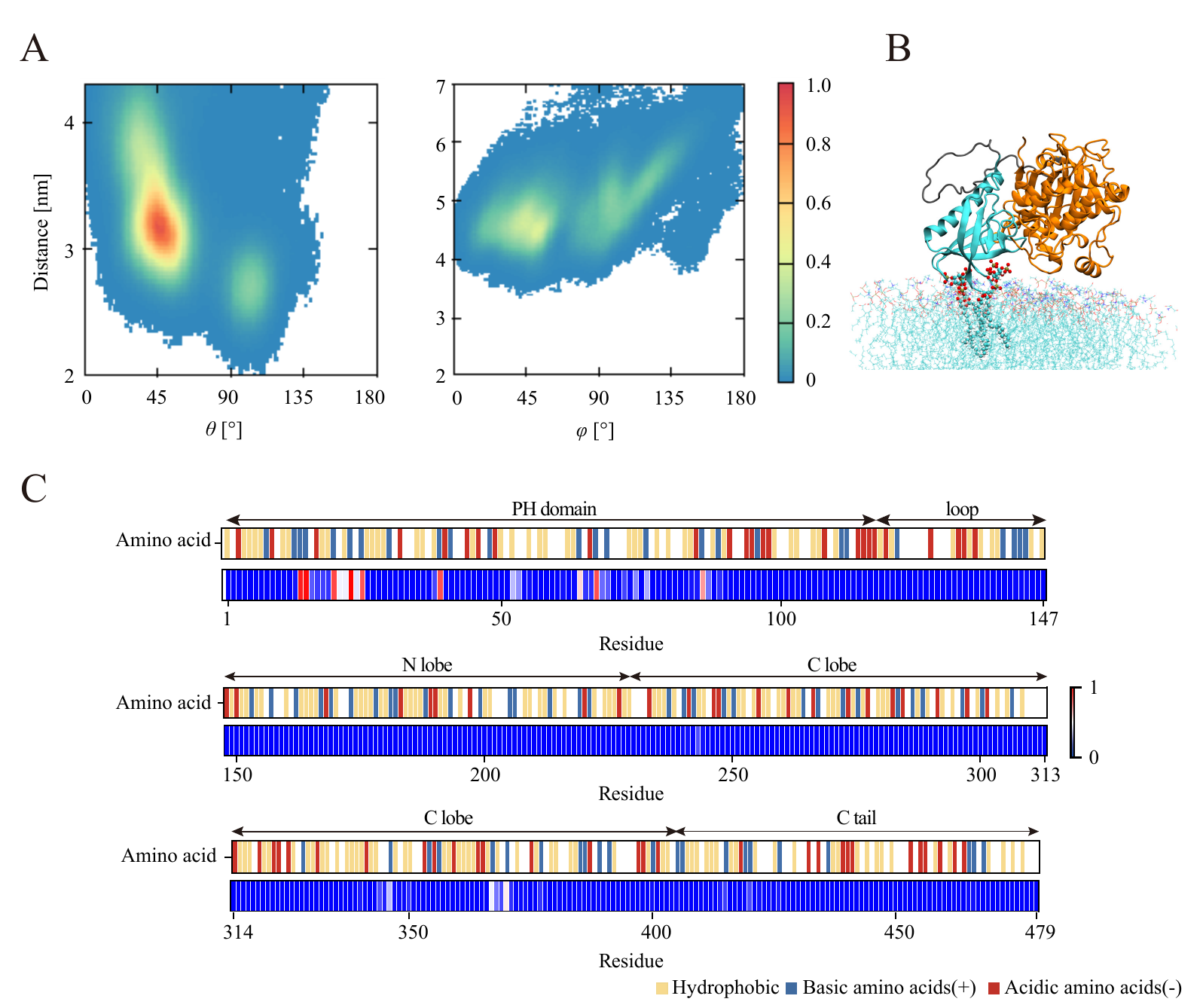}
\caption{AKT binding on a membrane with a reduced PIP$_3$ fraction.
(A)~Normalized density maps for the PH domain (left) and the kinase domain (right) shown as a function of the orientation angle and the $z$-component of the domain COM--bilayer distance.
(B)~Representative binding state of AKT.
(C)~Residues critical for PIP$_3$ association.
Normalized contacts were calculated using data from each cluster. For normalization, the number of contacts between a given residue and phosphate headgroups  was divided by the maximum number of contacts observed within each cluster.
Hydrophobic, basic, acidic, and other residues are shown in yellow, blue, red, and white, respectively.}
\label{fig5}
\end{figure*}

\subsection*{AKT adopts four membrane-binding modes on the lipid bilayer}
To investigate the orientation of AKT on the lipid bilayer, we measured the angles $\theta$ and $\phi$ between the bilayer normal and reference $\alpha$-helices in the PH and kinase domains, respectively (Fig.~\ref{orientation}A).
Using these angular data together with the distances between the COM of each domain and the lipid bilayer over time, we calculated two-dimensional probability density maps (Fig.~\ref{orientation}B and see Methods).
For the PH domain, we observed a strong peak at $\theta = 45\,\mathrm{\tcdegree}$ with a COM distance of $3.1$–$3.4\,\mathrm{nm}$, whereas the kinase domain exhibited a strong peak at $\phi = 90\,\mathrm{\tcdegree}$ with a COM distance of $5.1$–$5.7\,\mathrm{nm}$ (representative orientations are shown in Fig.~\ref{orientation}C).
Furthermore, these density maps indicate that both domains show several metastable orientations, which are represented by the lower-probability regions (yellow regions in Fig.~\ref{orientation}B).
These results reveal distinct preferred orientations for both domains, suggesting that AKT adopts several stable conformations and orientations on the lipid bilayer.

To further classify the binding modes of AKT on the membrane, we performed principal component analysis (PCA) followed by DBSCAN clustering (see Methods). 
We also calculated the values of $\theta$ and $\phi$ for each cluster(Fig.~S\ref{SI_PCA_angle}).
This analysis revealed four distinct binding modes of AKT on the lipid bilayer (Figs.~\ref{orientation}D, \ref{orientation}E).
We defined these four binding modes as Cluster I, Cluster \mbox{I\hspace{-1.2pt}I}, Cluster \mbox{I\hspace{-1.2pt}I\hspace{-1.2pt}I}, and Cluster \mbox{I\hspace{-1.2pt}V}, with $67\,\%$, $11\,\%$, $7\,\%$, and $3\,\%$ of membrane-bound configurations assigned to each cluster, respectively.
Note that both the protein orientations $\theta$ and $\phi$ (Fig.~S\ref{SI_PCA_angle}) and the interdomain distances (Fig.~S\ref{COM_PDF}) show clear separation consistent with the PCA-defined clusters.

In Cluster I, the PH domain adopts an upright orientation, with the N-lobe positioned away from the lipid bilayer and the C-lobe associated with the lipid bilayer.
The PH domain associates with both the N-lobe and the C-lobe of the kinase domain.
Because the $\alpha$-helix (T92--E116) in the PH domain is oriented away from the kinase domain (Fig.~\ref{orientation}E), interactions are instead observed between the kinase domain and the N-terminal region of the PH domain (Fig.~S\ref{domaininteraction}).

In Cluster \mbox{I\hspace{-1.2pt}I}, the PH domain adopts an orientation similar to that in Cluster I, whereas the kinase domain tilts toward the membrane, bringing both its N-lobe and C-lobe closer to the lipid bilayer.
In this cluster, the PH domain does not contact the N-lobe of the kinase domain (Fig.~S\ref{domaininteraction}).

In Cluster \mbox{I\hspace{-1.2pt}I\hspace{-1.2pt}I}, the PH domain lies roughly parallel to the membrane with its binding surface facing downward, and both the N-lobe and C-lobe of the kinase domain are close to the lipid bilayer.
This orientation of the PH domain corresponds to the most stable conformation obtained from simulations of the isolated PH domain~\cite{Yamamoto2016}.
Our simulations of full-length AKT thus reveal that the most populated orientation of the PH domain in the multidomain context differs from that observed for the PH domain alone.
The PH domain contacts both the N-lobe and C-lobe of the kinase domain, as in Cluster I.
In contrast to Cluster I, the $\alpha$-helix of the PH domain is oriented toward the kinase domain in Cluster \mbox{I\hspace{-1.2pt}I\hspace{-1.2pt}I} (Fig.~\ref{orientation}E), and therefore no interactions are observed between the kinase domain and the N-terminal region of the PH domain (Fig.~S\ref{domaininteraction}).

In Cluster \mbox{I\hspace{-1.2pt}V}, the PH domain also lies roughly parallel to the membrane but with its binding surface facing upward, and both the N-lobe and C-lobe of the kinase domain are close to the lipid bilayer.

In Cluster I, Cluster I\hspace{-1.2pt}I and Cluster \mbox{I\hspace{-1.2pt}V}, the phosphorylation sites of AKT (Thr308 and Ser473) in the kinase domain are exposed, thereby enabling other proteins to associate with and activate AKT.

\subsection*{Multiple PIP$_3$ molecules interact with distinct binding sites in each binding mode}
Because multiple PIP molecules are known to bind PMPs and regulate their interactions with membranes~\cite{Yamamoto2015, Yamamoto2016, Yamamoto2020, Soteriou2025}, we analyzed the key residues for PIP$_3$ binding and the local PIP$_3$ density around AKT.
A residue of AKT was considered to be in contact with PIP$_3$ if any heavy atom of the residue was located within $0.7\,\mathrm{nm}$ of any of the three phosphates of the lipid.
We observed that the contact residues are mainly positively charged residues, such as arginine and lysine, and that the combinations of contacting residues differ between binding modes (Fig.~\ref{contact}).
This preference for basic residues is consistent with previous studies on PMPs interacting with PIP lipids~\cite{Yamamoto2015, Yamamoto2016, Yamamoto2020, LeHuray2022, Soteriou2025}.
Furthermore, we observed that positively charged residues in the kinase domain also interact with PIP$_3$, although these contacts are more diffuse and transient than those in the PH domain consistent with the lower PIP$_3$ density observed for the kinase domain(Fig.~\ref{PIP_density}A).

To further characterize the mechanism of PIP$_3$ association, we analyzed the PIP$_3$ density around AKT for each binding mode (Fig.~\ref{PIP_density}A).
In Cluster I, the density of PIP$_3$ is high at both the canonical and secondary binding sites in the PH domain.
These binding sites are composed of many positively charged residues (Fig.~\ref{PIP_density}B), suggesting that electrostatic interactions contribute strongly to PIP$_3$ binding.
K14, R15, K20, R23, R25, and K39 are located in the canonical binding site, whereas K64, R67, and R86 are located in the secondary binding site(Fig.~\ref{PIP_density}C).
In addition, an arginine residue in the C-lobe of the kinase domain is observed to interact with PIP$_3$, indicating a secondary PIP$_3$-interacting patch on the kinase domain.
Unlike the PH domain, where basic residues form well-defined binding sites, the kinase domain presents a broader positively charged surface and engages PIP$_3$ more diffusely (Figs.~\ref{contact}, \ref{PIP_density}C).

In Cluster I\hspace{-1.2pt}I, the density of PIP$_3$ is also high at the canonical and secondary binding sites of the PH domain.
Furthermore, positively charged residues in the kinase domain interact with PIP$_3$.
Compared to Cluster I, the N-lobe of the kinase domain is positioned closer to the lipid bilayer (Fig.~\ref{orientation}E), but we do not observe direct PIP$_3$ contacts with the N-lobe.
As in Cluster I, PIP$_3$ interactions in the kinase domain are limited to residues forming the positively charged region of the C-lobe and remain weaker than those at the PH domain sites.

In Cluster \mbox{I\hspace{-1.2pt}I\hspace{-1.2pt}I}, the PH domain is located roughly parallel to the membrane with its binding surface facing downward, leading to PIP$_3$ contacts with a different residue (R76) compared to the upright orientation in Cluster I and Cluster I\hspace{-1.2pt}I.
This reflects the change in the relative orientation of the PH domain with respect to the bilayer.

In Cluster \mbox{I\hspace{-1.2pt}V}, the PIP$_3$-binding surface of the PH domain is oriented upward, resulting in unstable PIP$_3$ binding at the PH domain.
Consequently, PIP$_3$ interactions through the kinase domain become predominant, with positively charged residues in the kinase domain mainly interacting with PIP$_3$, albeit more weakly than the PH domain in the other binding modes.

\section*{Discussion}
In this study, using CG-MD simulations, we investigated the association of full-length AKT with PIP$_3$-containing lipid membranes.
Our simulations confirm that interactions between positively charged residues within the PH domain and PIP$_3$ are crucial for membrane association, as previously suggested from studies of the isolated PH domain~\cite{james1996specific,Yamamoto2016}.
In addition, we found that a positively charged surface region in the kinase domain can also interact with PIP$_3$, although these interactions are weaker and more diffuse than those in the PH domain.

We identified four distinct membrane-binding modes of AKT that differ in the orientation and membrane contacts of the PH and kinase domains.
In the most stable binding mode, PIP$_3$ engages both the canonical and secondary binding sites within the PH domain, while the kinase domain adopts an orientation in which the regulatory phosphorylation sites (Thr308 and Ser473) are exposed to the solvent.
Such a configuration is expected to facilitate efficient signal transduction, because it optimally presents these sites to upstream kinases such as PDK1 and mTORC2~\cite{Ebner2017}.

PIP$_3$ plays an essential role in various biological processes, including the localization of protein kinases, regulation of small GTPases, and tissue regeneration~\cite{Posor2022}.
Consequently, when multiple proteins interact with PIP$_3$ in the cellular membrane, they are likely to compete for a limited number of PIP$_3$ molecules.
As a result, AKT may experience an effective PIP$_3$ concentration lower than the bulk concentration in the membrane.
Previous work has shown that the orientation of PH domains on lipid bilayers can change depending on the concentration of PIP molecules~\cite{Yamamoto2020}.

To examine this effect for full-length AKT, we performed additional CG-MD simulations under conditions where the PIP$_3$ concentration was reduced so that only two PIP$_3$ molecules were present in each leaflet.
Under this condition, the peak corresponding to the stable orientation of the kinase domain observed at higher PIP$_3$ concentrations disappeared (Fig.~5A), suggesting that the kinase domain adopts less stably oriented configurations.
Furthermore, the AKT residues critical for PIP$_3$ binding were largely restricted to the PH domain, in contrast to the broader interactions involving the kinase domain observed at higher PIP$_3$ concentrations (Figs.~5B, 5C).
This indicates that, under PIP$_3$-limiting conditions, the PH domain effectively outcompetes the kinase domain for binding to the available PIP$_3$ molecules.
Specifically, at low PIP$_3$ concentration, interactions were confined to the most critical binding sites (canonical and secondary binding sites) in the PH domain, whereas at higher PIP$_3$ levels, additional stabilizing contacts were also observed, including those in the kinase domain.

Overall, our work highlights how the cooperative interplay between lipid recognition by PH domains and PIP$_3$-sensitive orientations of kinase domains governs the membrane binding of multidomain PMPs at PIP-containing membranes.
Cellular membranes are inherently heterogeneous, and the local concentration of PIP$_3$ can vary substantially across different membrane regions and cellular contexts.
Our findings suggest that such heterogeneous PIP$_3$ distributions can modulate both the strength and the mode of protein–membrane interactions, as exemplified here by AKT, and thereby influence the efficiency and outcome of signal transduction.
The CG-MD framework established here can be readily extended to other multidomain PMPs to investigate whether similar PIP-dependent switching between binding modes and activation-competent orientations may represent a general principle of membrane-associated signaling.

\section*{Methods}

\subsection*{Rescaling protein--protein interactions}
For the full-length AKT structure, we used an AlphaFold2-predicted model (AlphaFold DB: AF-P31749-2-F1~\cite{jumper2021highly,varadi2022alphafold}), in which the PH domain was replaced by the crystal structure (PDB ID: 1UNQ~\cite{Milburn2003}) because it contains a PIP$_3$ headgroup bound at the canonical binding site.
We confirmed that the kinase domain in the AlphaFold-predicted model is in good agreement with the crystal structure (PDB ID: 4GV1~\cite{addie2013discovery}) (Fig.~S\ref{support1}B).

Coarse-grained molecular dynamics (CG-MD) simulations were performed using GROMACS 2022.6~\cite{Abraham2015} and the Martini 2.2 force field~\cite{Jong2013}.
Because the Martini force field is known to overestimate protein--protein interactions~\cite{Stark2013, Thomasen2022a}, we adjusted the force field by weakening the Lennard--Jones (LJ) potentials involved in protein--protein interactions~\cite{Thomasen2024}.
To estimate an appropriate scaling factor, we first performed CG-MD simulations of isolated AKT in solution (Fig.~S\ref{support1}C).
The simulation box size was set to $20 \times 20 \times 20 \,\mathrm{nm}^3$, which is large enough to avoid self-interactions across periodic boundaries.
The system was solvated with approximately 63,000 CG water beads, and NaCl ions were added to neutralize the system at a concentration of $150\,\mathrm{mM}$.
After 1{,}000 steps of energy minimization, all systems were equilibrated for $1\,\mathrm{ns}$ at 310\,K and 1\,bar using the Berendsen thermostat and barostat~\cite{Berendsen1984}.

Simulations were performed under three different LJ scaling conditions, where the protein--protein LJ interaction strength was scaled to $91\,\%$, $92\,\%$, or $93\,\%$ of the standard Martini 2.2 parameters.
For each condition, three independent production simulations were carried out for $10\,\mu\mathrm{s}$ each (i.e., $3 \times 10\,\mu\mathrm{s}$ per scaling factor) with a time step of $20\,\mathrm{fs}$.
To maintain the secondary structure of each domain, an elastic network model was applied with a cutoff range of $0.5$--$0.9\,\mathrm{nm}$~\cite{Atilgan2001}.
The temperature was maintained at $310\,\mathrm{K}$ using the velocity-rescaling thermostat~\cite{Bussi2007}, and the pressure was kept at $1\,\mathrm{bar}$ using the Parrinello--Rahman barostat~\cite{Parrinello1981}.
The bond lengths were constrained to equilibrium lengths using the LINCS algorithm~\cite{Hess1997}. Lennard-Jones and Coulombic interactions are cut off at $1.1 \,\mathrm{nm}$, with the potentials shifted to zero at the cutoff.

To quantify conformational changes of AKT in solution, we calculated the radius of gyration ($R_g$).
Under all tested conditions, $R_g$ converged to smaller values, indicating a transition to a more compact structure (Fig.~S\ref{support1}D).
The time-averaged $R_g$ values were $2.49 \pm 0.15\,\mathrm{nm}$ (91\%), $2.54 \pm 0.17\,\mathrm{nm}$ (92\%), and $2.50 \pm 0.15\,\mathrm{nm}$ (93\%).
These values are in close agreement with the previously reported $R_g$ for full-length AKT ($2.63$--$2.70\,\mathrm{nm}$)~\cite{Lucic2018}.

Furthermore, we validated the scaling factor using simulations of the intrinsically disordered protein ACTR.
The $R_g$ of ACTR in all-atom MD simulations has been reported to be $2.2$--$2.3\,\mathrm{nm}$~\cite{Zheng2018}, and this value was reproduced when the protein--protein interaction strength was scaled down to 93\% (Fig.~S1A).
Based on these results, we adopted a 93\% scaling of protein--protein LJ interactions for the Martini 2.2 force field in all subsequent simulations.

\subsection*{Simulations of AKT with a lipid bilayer}
CG-MD simulations of full-length AKT interacting with a lipid bilayer were performed using the Martini 2.2 force field, with protein--protein LJ interactions scaled to 93\%.
The symmetric lipid bilayer was constructed using CHARMM-GUI~\cite{Jo2008} and consisted of 632 POPC (79\%), 152 POPS (19\%), and 16 PIP$_3$ (2\%) molecules.
The simulation box size was approximately $15 \times 15 \times 30\,\mathrm{nm}^3$.
Non-phosphorylated AKT (Thr450 unphosphorylated) was initially placed $15\,\mathrm{nm}$ above the membrane.
Each system was solvated with approximately 52,000 CG water beads, and NaCl ions were added to neutralize the system at a concentration of $150\,\mathrm{mM}$.
After 1,000 steps of energy minimization, all systems were equilibrated at constant $NPT$ for $1\,\mathrm{ns}$.
Production simulations were performed for $10 \times 50\,\mathrm{\mu s}$, starting from different initial orientations of AKT.

\subsection*{Orientation Angles and Density Maps}
The angle $\theta$ was defined as the angle between the membrane normal ($z$-axis) and the $\alpha$-helix formed by residues T92--E116 in the PH domain of AKT (Fig.~\ref{orientation}A).
Similarly, the angle $\phi$ was defined as the angle between the $z$-axis and the $\alpha$-helix formed by residues R328--C344 in the kinase domain (Fig.~\ref{orientation}A).

AKT was defined as membrane-bound when the minimum distance between any residue in the PH domain and the lipid bilayer (defined by the COM of the phosphate beads) was less than $4.3\,\mathrm{nm}$. 
For membrane-bound AKT, two-dimensional probability density maps were constructed for each domain in the $(\theta, d)$ and $(\phi, d)$ spaces, where $d$ denotes the $z$-component of the distance between the domain COM and the lipid bilayer COM.

Principal component analysis (PCA)~\cite{Wold1987} followed by DBSCAN clustering~\cite{Ester1996} was applied to angular data extracted every $100\,\mathrm{ns}$ from all $10 \times 50\,\mathrm{\mu s}$ simulations.
The analysis was restricted to membrane-bound configurations.
The four-dimensional data set, consisting of $\theta$, $\phi$, and two additional angles defined for the PH domain (K64$\rightarrow$C60) and the kinase domain (E247$\rightarrow$K268), was reduced to two dimensions using PCA, and the resulting data were clustered using DBSCAN.
A cluster was defined when at least 50 points were found within a radius of 0.3 in the PCA space.

\subsection*{Atomistic reconstruction and visualization}
Representative coarse-grained structures obtained from the simulations were converted to all-atom models using CG2AT~\cite{Vickery2021}.
Simulation trajectories were visualized using VMD~\cite{Humphrey1996}.
Electrostatic potential maps were calculated using PDB2PQR~\cite{Dolinsky2004} and the Adaptive Poisson--Boltzmann Solver (APBS)~\cite{Baker2001}.

\subsection*{Acknowledgments}
This work was supported by JSPS KAKENHI Grant Number 22K06171, Japan.

%


\clearpage
\onecolumngrid
\clearpage

\setcounter{page}{1}
\pagenumbering{arabic} 

{
    \center \bf \large l
    Supplemental information\vspace*{1cm}\\ 
    \vspace*{0.0cm}
}

\setcounter{figure}{0}

\begin{figure*}[h]
\centering
\includegraphics[width = 145mm,bb= 0 0 930 1186]{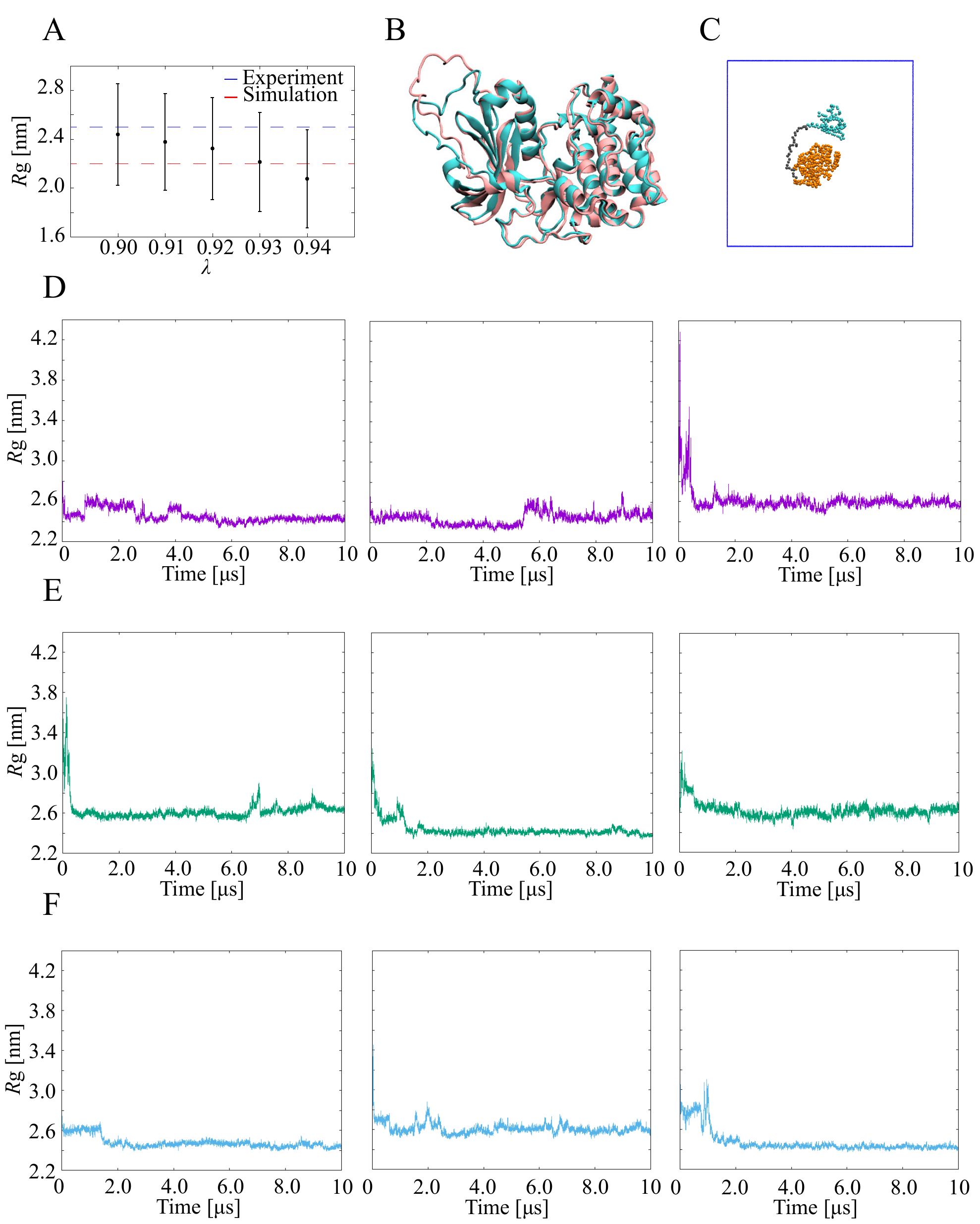}
\caption{Simulations for rescaling protein--protein interactions to reproduce the compactness of full-length AKT. 
(A)~Relationship between the protein--protein interaction scaling parameter $\lambda$ and the radius of gyration $R_g$. 
(B)~Structural comparison between the crystal structure (PDB ID: 4GV1, cyan) and the AlphaFold2-predicted model (pink). 
(C)~Simulation system used for the rescaling procedure. 
(D--F)~Time series of $R_g$ at $\lambda = 91\%$ (D), $92\%$ (E), and $93\%$ (F).}
\label{support1}
\end{figure*}

\begin{figure*}[tb]
\centering
\includegraphics[width = 150mm,bb= 0 0 564 250]{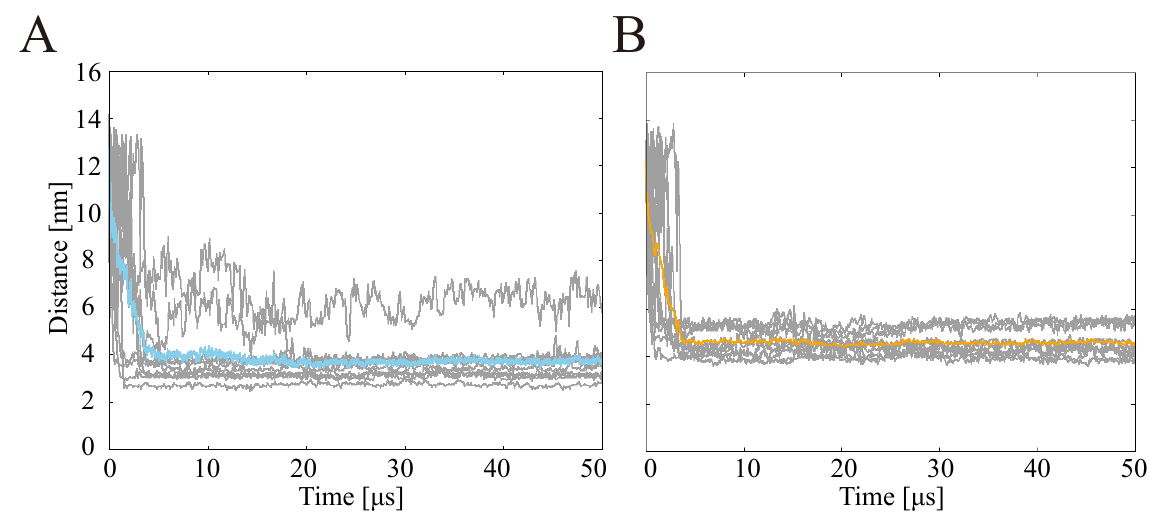}
\caption{Time series of the distance between the domain COM and the lipid bilayer.
(A)~PH domain. (B)~Kinase domain.}
\label{support2}
\end{figure*}

\begin{figure*}[tb]
\centering
\includegraphics[width = 80mm,bb= 0 0 546 541]{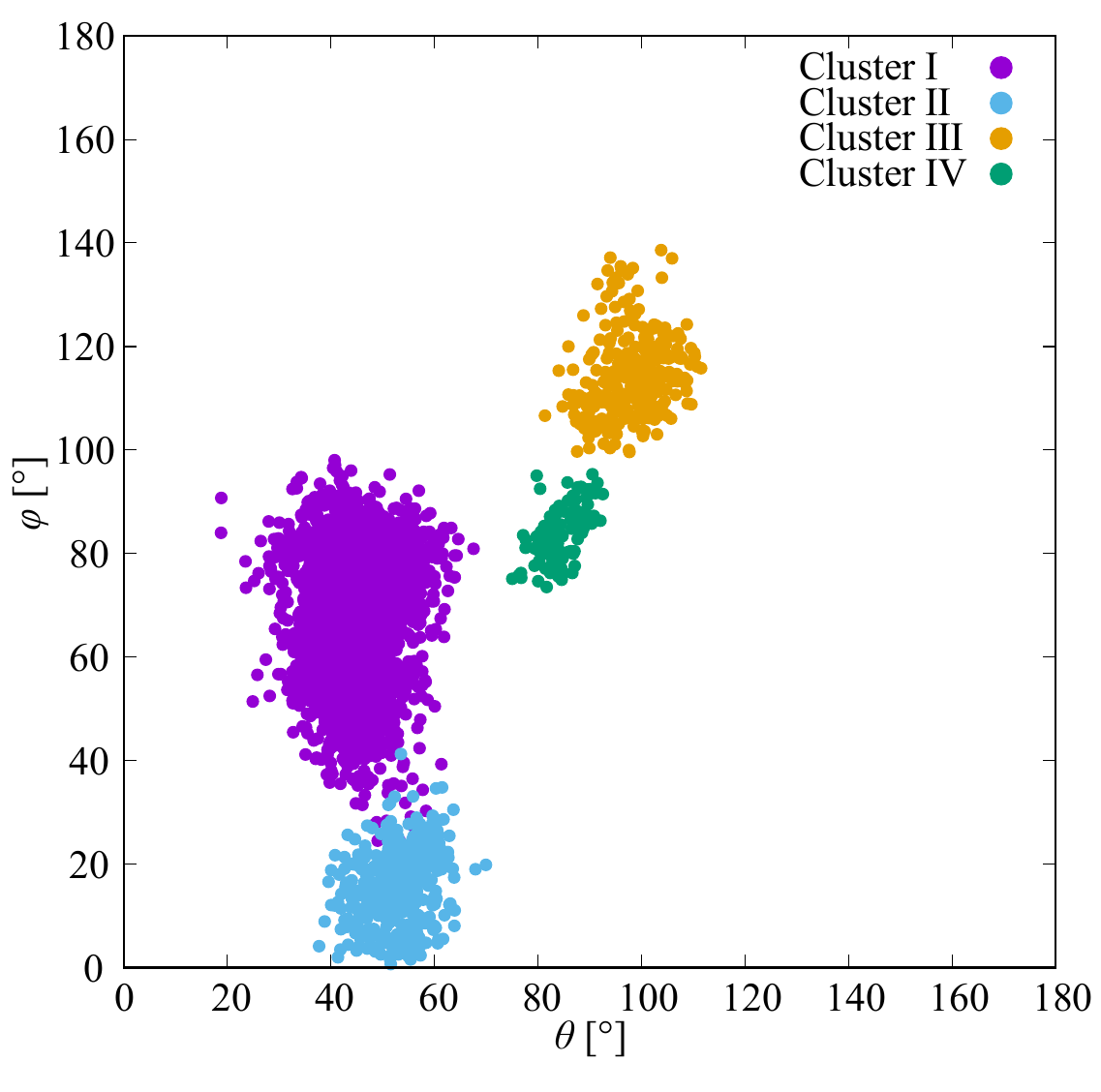}
\caption{Distribution of $\theta$ and $\phi$ for membrane-bound AKT.
Data points are colored according to the binding modes identified by PCA followed by DBSCAN clustering.
The data clearly separate into four distinct clusters.}
\label{SI_PCA_angle}
\end{figure*}

\begin{figure*}[tb]
\centering
\includegraphics[width = 85 mm,bb= 0 0 503 383]{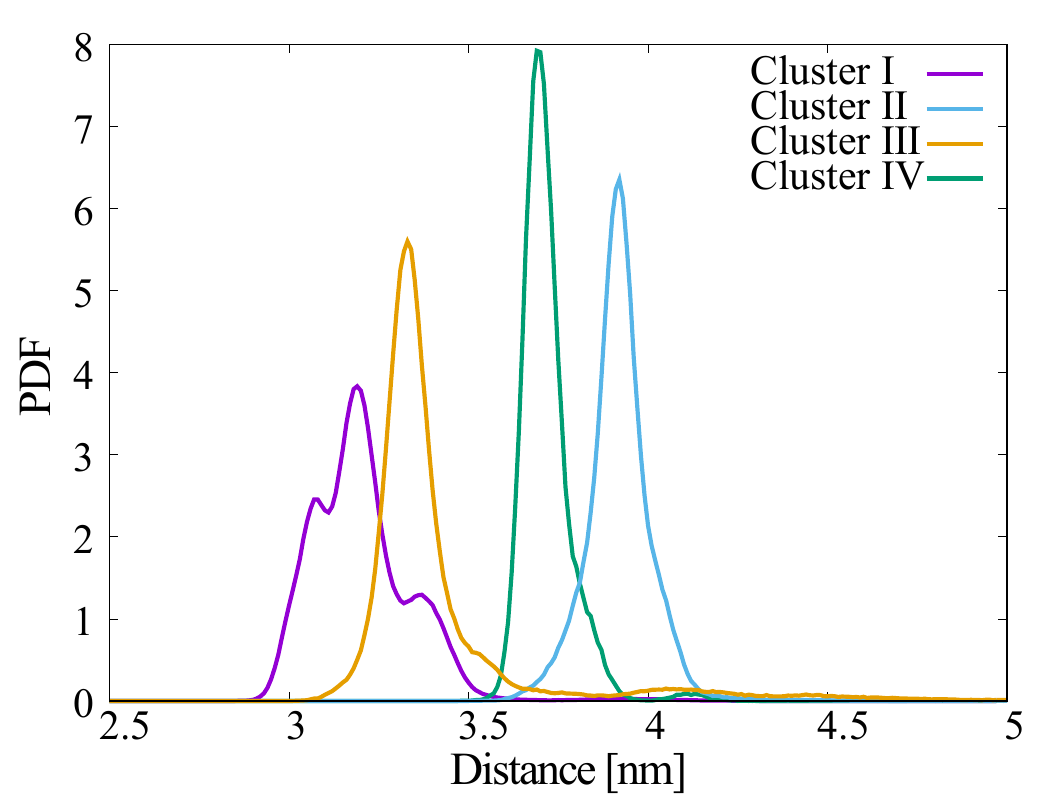}
\caption{Probability density functions of the intramolecular distance between the COM of the PH and kinase domains for each cluster.
Peak positions differ among clusters.
Multiple peaks in Cluster I reflect slight differences in membrane-associated compaction rather than distinct structural states.}
\label{COM_PDF}
\end{figure*}

\begin{figure*}[tb]
\includegraphics[width = 130 mm, bb= 0 0 538 599]{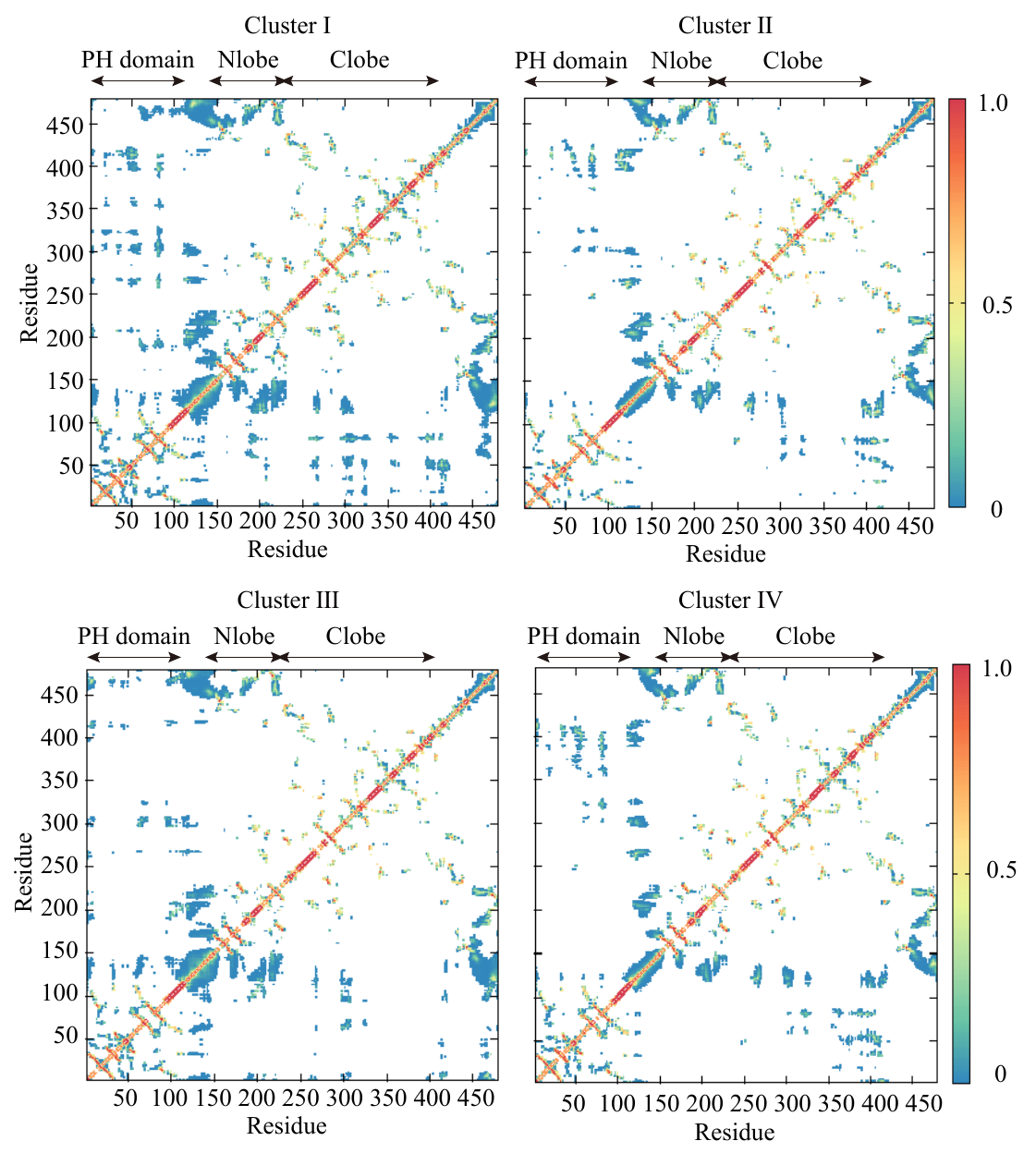}
\caption{Intramolecular contacts of AKT in each binding mode.}
\label{domaininteraction}
\end{figure*}

\end{document}